\documentstyle[nato]{crckapb}
\begin{opening}
\title{Abelian and Nonabelian Lattice Chiral Gauge Theories 
through Gauge Fixing}
\author{Wolfgang Bock}
\author{Ka Chun Leung}
\institute{Institute of Physics, University of Siegen\\
57068 Siegen, Germany}
\author{Maarten Golterman}
\institute{Department of Physics, Washington University\\
St. Louis, MO 63130, USA}
\author{Yigal Shamir}
\institute{Beverly and Raymond Sackler Faculty of Exact Sciences\\
Tel-Aviv University, Ramat Aviv 69978, Israel}
\end{opening}
\runningtitle{Lattice Chiral Gauge Theories}
\begin{document}
\newcommand{\psibar}{{\overline{\psi}}}
\newcommand{\Dsl}{D\!\!\!\!/}
\newcommand{\dsl}{\partial\!\!\!/}
\begin{abstract}
After an introduction in which we review the fundamental 
difficulty in constructing lattice chiral gauge theories,
we summarize the analytic and numerical evidence that abelian
lattice chiral gauge theories can be nonperturbatively constructed
through the gauge-fixing approach.  In addition, we indicate how
we believe that the method may be extended to nonabelian chiral
gauge theories.
\end{abstract}
\section{Introduction}
Consider a collection of left-handed fermion fields transforming in a
representation of some symmetry group.  A gauge theory containing
these fermions can be regulated by putting it on a (euclidean)
space-time lattice.  We may then ask about the anomaly structure
of the theory by keeping the gauge fields external (and smooth),
and it is clear that each fermion field will have to contribute
its share to the expected chiral anomaly.  This can happen in
two ways: either the regulated theory is exactly invariant under
the full symmetry group, and each fermion comes with its species
doublers \cite{ks,nn}, or the symmetry is somehow explicitly broken
by the regulator (in this case the lattice), making it possible
for the fermion field in each irreducible representation 
to produce the correct contribution to the anomaly in the continuum limit 
({\it i.e.} for smooth external gauge fields).

The Nielsen-Ninomiya theorem \cite{nn} tells us that fermion
representations with doublers contain equally many left- and right-handed
fermions transforming the same way under the symmetry group.  This
implies that if we make the gauge fields dynamical, a vector-like
gauge theory will emerge.  Hence, if we wish to construct a genuinely
chiral theory on the lattice, we will have to resort to the
second option: an explicit breaking, by the lattice regulator, of
the symmetry group.  The most well-known example of this is the
formulation of lattice QCD with Wilson fermions \cite{wilson}.
In this method, a momentum-dependent mass term of the form
\begin{equation}
-\frac{r}{2}\sum_\mu\left(\psibar_x\psi_{x+\mu}+\psibar_{x+\mu}\psi_{x}
-2\psibar_x\psi_x\right)  \label{WMASS}
\end{equation} 
is added to the action, which removes the doublers by giving them a 
mass of the order of the cutoff $1/a$ (where $a$ is the lattice spacing,
which we set equal to one in this talk). 

For theories in which only vector-like symmetries are gauged,
like QCD, this works fine.  The Wilson mass
term can be made gauge invariant (by inserting the SU(3)-color link
variables on each hopping term).  The global chiral symmetry is broken,
but can be restored in the continuum limit by a subtraction of the quark
mass.

However, the situation changes dramatically when we wish to gauge a 
chiral symmetry.  We can still remove the doublers 
with a Wilson mass term as in Eq.~(\ref{WMASS}), by introducing a
right-handed ``spectator" fermion for each left-handed fermion.
(Other possibilities exist, but, because of the anomaly argument,
the conclusions are similar in all cases \cite{ysnogo}.)  Obviously, since we are
now interested in gauging a chiral symmetry, the Wilson mass term
does not respect gauge invariance.  This means that, on the lattice,
the longitudinal gauge field (which represents the gauge degrees
of freedom) couples to the fermions.  If we only have a Maxwell-like
term ($\sim{\rm tr}\;F_{\mu\nu}^2$) controlling the dynamics of the gauge
field, the longitudinal modes are not suppressed at all, and they
typically destroy the chiral nature of the fermion spectrum
(see Refs.~\cite{ys,dp} for reviews and references).  Note that 
this phenomenon is
nonperturbative in nature.  The problem is invisible for smooth
gauge fields, but the point is that longitudinal gauge fields do not
have to be smooth, even for small gauge coupling, if all gauge fields
on an orbit have equal weight in the partition function. 

This is precisely where gauge fixing comes in.  A renormalizable
choice of gauge adds a term to the gauge-field action which controls
the longitudinal part of the gauge field.  In this talk, we will
consider the Lorentz gauge, with gauge-fixing lagrangian
$(1/2\xi){\rm tr}\;(\partial_\mu A_\mu)^2$.  The longitudinal
part of the gauge field ($\partial_\mu A_\mu)$ has now acquired the
same ``status" as the transverse part ($F_{\mu\nu}$). 

Before we start the discussion of gauge fixing, it is instructive
to see in more detail what goes wrong without it, using our example
of a Wilson mass term.  If we perform a gauge transformation on
the left-handed fermion field, $\psi_L\to\phi^\dagger\psi_L$, with
$\phi$ a group-valued scalar field, the Wilson mass term transforms
into
\begin{equation}
-\frac{r}{2}\sum_\mu\left(\psibar_{Rx}\phi^\dagger_{x+\mu}\psi_{Lx+\mu}
+\dots\right)\;. \label{WY}
\end{equation}
The parameter $r$ is promoted to a Yukawa-like coupling,
and the lattice regulator (which led to the introduction of the Wilson mass
term in the first place) leads to couplings between the fermions and
the longitudinal degrees of freedom through the scalar field $\phi$.
Note that the lattice theory {\it is} invariant under the symmetry
$\psi_{L,R}\to h_{L,R}\psi_{L,R}$, $\phi\to h_L\phi h^\dagger_R$
{\it etc.}, with $h_R$ global and $h_L$ local ($h$-symmetry).  
The $h_L$-symmetry is, however, not
the same as that of the gauge theory we wish to construct, since $\phi$
is supposed to decouple.

We can now explore the phase diagram ({\it i.e.} all values of $r$)
in order to see whether we might decouple these longitudinal modes, while
retaining the fermion spectrum that we put in.  (In a confining theory,
of course the ``chiral quarks" do not appear in the spectrum, but we can
imagine first considering the theory with only the gauge degrees $\phi$
dynamical, with external smooth transverse gauge fields.  This is the
so-called ``reduced model" of Refs.~\cite{wmypd,wmyprl,wmylat97}.)

It turns out that three things can happen (see Ref.~\cite{dp} and 
refs. therein).
The $h$-symmetry can be spontaneously broken, and the doublers will
be removed if $\langle\phi\rangle\sim 1/a$.  However, in that case
also the gauge-field mass will be of order $1/a$, which is not what we
want.  It follows that we would like the $h$-symmetry to be unbroken.

For small $r$, we may read off the fermion spectrum by replacing
$\phi\to\langle\phi\rangle$.  If $\langle\phi\rangle=0$, we find that
the Wilson-Yukawa term does not lead to any fermion masses, and the
doublers are degenerate with the massless physical fermion!  (In the
broken phase, this degeneracy is partially removed, but, as we already
noticed, the doubler masses will be set by the scale of the gauge-field
mass.  An elegant way of doing this was reviewed in Ref.~\cite{montvay}.)
There also exists a phase with unbroken $h$-symmetry at large $r$,
and it turns out that in that phase the only massless left-handed fermion
is described by the composite field $\phi^\dagger\psi_L$.  This
fermion, however, does not couple to the gauge field, since its
gauge charge is ``screened" by the longitudinal field $\phi$
\cite{mdj,mdrome}.  Both this composite left-handed and the spectator
right-handed fermion do not couple to the gauge field (in four
dimensions; for two dimensions, see Ref.~\cite{twod}).

What we will show in the rest of this talk is that this conclusion,
that there is no place in the phase diagram where a chiral gauge theory
can be defined, changes completely when  a gauge-fixing term is added,
therewith enlarging the parameter space of the phase diagram.
We mention here that another approach exists which aims to ``tame 
the rough gauge fields" (interpolation, or two-cutoff approach), see
Ref.~\cite{ys} for a review and references. 

Before we end this introduction, 
we would like to rephrase our conclusions thus far in a 
somewhat different way.  Imagine that we have defined the fermionic
partition function for an external lattice gauge field (not 
necessarily smooth!) in a certain
attempt to construct a chiral gauge theory.   This then yields
an effective action $S_{\rm effective}(A)$, where $A$ is the external 
gauge field, and we have, under a gauge transformation,
\begin{equation}
\delta S_{\rm effective}(A)={\rm anomaly}(A)+{\rm lattice\ artifacts}(A)\;.
\label{SEFFECTIVE}
\end{equation}
The anomaly part can be identified by 
choosing the external gauge field to be small and smooth.
The lattice-artifact terms are generically not small.  They cannot
be, if the dynamics of the longitudinal modes is to change the
theory into a vector-like theory (as described above), in which the
gauge anomaly vanishes.  This points to
another way of avoiding the conclusions sketched above: by finding a
fermion partition function for which the lattice-artifact
term vanishes, if the fermion representation is anomaly-free in the
continuum, for {\it all} lattice gauge fields.  
A recent proposal along these lines, starting from
a Dirac operator satisfying the Ginsparg--Wilson relation, is
reviewed in Ref.~\cite{luescher}. 

\section{Gauge fixing -- the abelian case}

The central idea of the gauge-fixing approach is to make gauge fixing
part of the definition of the theory \cite{rome}.  This contrasts with the case
of lattice QCD, where, because of the compact nature of the lattice
gauge fields, gauge fixing is not needed.  The theory is defined by
the action
\begin{equation}
S=S_{\rm gauge}+S_{\rm fermion}+S_{\rm g.f.}+S_{\rm ghosts}+S_{\rm c.t.}\;.
\label{ACTION}
\end{equation}
For $S_{\rm gauge}$ we choose the standard plaquette term.  For 
$S_{\rm fermion}$ we use Wilson fermions, with only the left-handed
fermions coupled to the gauge fields.  We will choose the Wilson mass
term as in Eq.~(\ref{WMASS}), without any gauge fields in the hopping
terms.  Other choices are possible, but in a chiral gauge theory, all
break the gauge symmetry.  Our choice has the (technical) advantage of
making the action invariant under shift symmetry, $\psi_R\to\psi_R
+\epsilon_R$, with $\epsilon_R$ a constant, right-handed Grassmann
spinor \cite{gp}.  For $S_{\rm g.f.}$ we will choose a lattice
discretization of $\int d^4x (1/(2\xi))(\partial_\mu A_\mu)^2$, to be
discussed below.  In this section, we will be concerned only with
abelian theories, and $S_{\rm ghosts}$ can be omitted \cite{wmynoghosts}.

Since the lattice regulator breaks gauge invariance explicitly, counterterms
are needed, and they are added through $S_{\rm c.t.}$.  These counterterms
include one dimension 2 operator (the gauge-field mass counterterm),
no dimension 3 operators (because of the shift symmetry), and a host of
dimension-4 counterterms (see Refs.~\cite{rome,wmky} for a detailed discussion).
Tuning these counterterms to the appropriate values (by requiring the
Slavnov--Taylor identities of the continuum target theory to be
satisfied) should then bring us
to the critical point(s) in the phase diagram at which a chiral gauge
theory can be defined.  Because of the choice of a renormalizable gauge,
it is clear that this can be done in perturbation theory (if the theory
is anomaly free).  The observation of Ref.~\cite{rome} is that also
nonperturbatively gauge-fixing will be needed in order to make the
program described above work.

At the nonperturbative level, the following important questions arise
\cite{ysrome}.  First, what should we choose as the lattice discretization
of $S_{\rm g.f.}$?  More precisely, given a certain choice, what does
the phase diagram look like, and for which choices do we find a phase
diagram with the desired critical behavior?  Second, if we find that a
suitable discretization exists, so that the fermion content is indeed
chiral, how does this precisely happen?  Note that, without gauge fixing,
the action above is essentially just the Smit--Swift model \cite{ssm},
which, for the reasons summarized in the introduction, does not work.
In this section, we will answer these two questions.  We relegate the
discussion of a third important question, namely the extension to
nonabelian theories, to section 3.

\subsubsection*{Gauge fixing on the lattice}
\medskip
It was argued in Ref.~\cite{ysrome} that the lattice gauge-fixing term 
\begin{equation}
S_{\rm g.f., naive}=\frac{1}{2\xi g^2}\sum_x\left(
\sum_\mu({\rm Im}\;U_{x,\mu}-{\rm Im}\;U_{x-\mu,\mu})\right)^2
\label{NAIVE}
\end{equation}
is {\it not} the right choice, even though, expanding the link
variables $U_{x,\mu}={\rm exp}(igA_{x,\mu})$, it looks like the most
straightforward discretization of the continuum form.  This is because
this choice admits an infinite set of lattice Gribov copies (which
have no continuum counterpart) of the perturbative vacuum $U_{x,\mu}=1$.
This is dangerous, since our intuition that this approach should lead
us to the lattice construction of chiral gauge theories is
based on the fact that our regulator does work in perturbation theory.
Therefore, we insist that lattice perturbation theory should be a 
reliable approximation of our
lattice theory at weak coupling.  In fact, we showed, through a
combination of numerical and mean field techniques, that the naive
choice of gauge-fixing action of Eq.~(\ref{NAIVE}) does not lead to
a phase diagram with the desired properties \cite{wmky}.

The vacuum degeneracy of $S_{\rm g.f., naive}$ can be lifted by adding
irrelevant terms to it \cite{ysrome,my}, so that
\begin{equation}
S_{\rm g.f.}=S_{\rm g.f., naive}+{\tilde r}S_{\rm irrelevant}\;,
\label{GAUGEFIX}
\end{equation}
where $\tilde r$ is a parameter very similar to the Wilson parameter
$r$ multiplying the Wilson mass term.  While we will not give any
explicit form of $S_{\rm irrelevant}$ here, it was shown \cite{my}
that $S_{\rm irrelevant}$ can be chosen such that
\begin{equation}
S_{\rm g.f.}(U)\ge 0\ \ \ {\rm and}\ \ \ S_{\rm g.f.}(U)=0
\Leftrightarrow U_{x,\mu}=1\;.\label{PROP}
\end{equation}
This means that $U_{x,\mu}=1$ is the unique perturbative vacuum.
Also, obviously, $S_{\rm g.f.}(U)\to \int d^4x (1/2\xi)(\partial_\mu
A_\mu)^2$ in the classical continuum limit.  Our choice does not
respect BRST symmetry, which will necessitate adjustment of the
counterterms \cite{wmynoghosts}.

For small gauge coupling $g$, the classical potential should give us
an idea of what the phase diagram looks like.  Without fermions
(which contribute to the gauge-field effective potential only at
higher orders in lattice perturbation theory), including (only) a 
mass counterterm $-\kappa\sum_\mu(U_{x,\mu}+U^\dagger_{x,\mu})$, and
expanding $U_{x,\mu}={\rm exp}(igA_{x,\mu})$, we have, for our choice of
$S_{\rm irrelevant}$,
\begin{equation}
V_{\rm classical}(A)=\frac{{\tilde r}g^4}{4\xi}\left(\sum_\mu A^2_\mu
\sum_\nu A^4_\nu+\dots\right)+\kappa g^2\left(\sum_\mu A^2_\mu+\dots
\right)
\label{CLASS}
\end{equation}
for a constant field.  The dots indicate higher-order terms in $g^2$.
While the precise form of the term proportional to $\tilde r$ is not
important, it is clearly irrelevant (order $A^6$) and positive
({\it i.e.} it stabilizes the perturbative vacuum).  
 
We can now distinguish two different phases, depending on the value
of $\kappa$.  For $\kappa>0$, $A_\mu=0$, and the gauge field has a
positive mass $\sqrt{2\kappa g^2}$.  For $\kappa<0$, the gauge field
acquires an expectation value $A_\mu=\pm\left(-\frac{\xi\kappa}
{3{\tilde r}g^2}\right)^{1/4}$, for all $\mu$, and we encounter a 
new phase, in which the (hypercubic) rotational symmetry is 
spontaneously broken!
These two phases are separated by a continuous phase transition
(classically at $\kappa=\kappa_c=0$), at which the gauge-field mass
vanishes.  In other words, we will take our continuum limit by
tuning $\kappa\searrow\kappa_c$.  (For a discussion including the
other, dimension 4, counterterms, see Ref.~\cite{my}.)

A detailed analysis of the phase diagram for the abelian
theory without fermions was given in Ref.~\cite{wmky}.  A complete
description of the phase diagram in the four-parameter space
spanned by the couplings $g$, $\xi$, $\tilde r$ and $\kappa$ can be
found there, as well as a discussion of the other counterterms and
a study of gauge-field propagators.
In the region of interest (basically small $g$ and $\tilde r\approx 1$)
good agreement was found between a high-statistics numerical study and
lattice perturbation theory.  In particular, the picture that emerges
from the classical potential as described above was shown to be
correct, as long as we choose $\tilde r$ away from zero,
and $g^2$, $\xi g^2$ sufficiently small.  As it should, 
the theory (without fermions) at the critical point describes free, 
relativistic photons.

\subsubsection*{Fermions}
\medskip
We now come to the behavior of the fermions in this gauge-fixed lattice
theory.  Employing a continuum-like notation for simplicity, our lattice
lagrangian, including fermions, reads
\begin{eqnarray}
{\cal L}&=&\frac{1}{4}F_{\mu\nu}^2+\frac{1}{2\xi}(\partial_\mu A_\mu)^2
+{\tilde r}{\cal L}_{\rm irrelevant}(A) \label{VECTOR} \\
&&+\psibar\left(\Dsl(A)P_L+\dsl P_R\right)\psi-\frac{r}{2}\psibar\Box\psi
\nonumber \\
&&+\kappa g^2 A_\mu^2+\mbox{other counterterms}\;. \nonumber
\end{eqnarray}
In order to investigate the interaction between fermions and longitudinal
modes, we can make the latter explicit by a gauge transformation
\begin{eqnarray}
A_\mu&\to&\phi^\dagger A_\mu\phi-\frac{i}{g}\phi^\dagger\partial_\mu\phi
\equiv -\frac{i}{g}\phi^\dagger D_\mu\phi\;,
\label{GAUGETR} \\
\psi_L&\to&\phi^\dagger\psi_L\;.\nonumber
\end{eqnarray}
This yields the lagrangian in the ``Higgs" or ``St\"uckelberg" picture,
\begin{eqnarray}
{\cal L}&=&\frac{1}{4}F_{\mu\nu}^2+\frac{1}{2\xi g^2}\left(\partial_\mu 
(\phi^\dagger(-i\partial_\mu+gA_\mu)\phi)\right)^2
+{\tilde r}{\cal L}_{\rm irrelevant}(A,\phi) \label{HIGGS} \\
&&+\psibar\left(\Dsl(A)P_L+\dsl P_R\right)\psi-\frac{r}{2}\left(\psibar_R
\Box(\phi^\dagger\psi_L)+\psibar_L\phi\Box\psi_R\right)
\nonumber \\
&&+\kappa\left(D_\mu(A)\phi\right)^\dagger
\left(D_\mu(A)\phi\right)+\mbox{other counterterms}\;. \nonumber
\end{eqnarray}
This action is invariant under the $h$-symmetry mentioned in the 
introduction.  

In order to find out whether the longitudinal modes, which are 
represented by the field $\phi$ in the Higgs-picture lagrangian,
change the fermion spectrum, we may now simplify the theory, by
first considering the ``reduced" model, in which we set $A_\mu=0$
in Eq.~(\ref{HIGGS}).  Expanding $\phi={\rm exp}(i\sqrt{\xi}g\theta)$,
which is appropriate for small gauge coupling because of the $1/g^2$
in front of the gauge-fixing term, gives the reduced-model lagrangian
\begin{eqnarray}
{\cal L}_{\rm red}
&=&\frac{1}{2}(\Box\theta)^2+\kappa\xi g^2(\partial_\mu\theta)^2
+\theta\ \mbox{self-interactions} \label{REDUCED} \\
&&+\psibar\dsl\psi-\frac{r}{2}\psi\Box\psi
+i\sqrt{\xi}g(\psibar_L\theta\Box\psi_R-\psibar_R\Box(\theta\psi_L))+
O(g^2)\;. \nonumber
\end{eqnarray}
This lagrangian teaches us the following.  First, $\theta$ is
a real scalar field with dimension 0, and inverse propagator
$p^2(p^2+2\kappa\xi g^2)$.  Near the critical point (which is
at $\kappa=0$ to lowest order), this behaves like $p^4$.  This 
actually implies \cite{wmypd,wmypt} that 
\begin{equation}
\langle\phi\rangle\propto (\kappa-\kappa_c)^{\xi g^2/(32\pi^2)}\to 0
\label{VEV}
\end{equation}
for $\kappa\to\kappa_c$.  (This behavior is very similar to that of
a normal scalar field in two dimensions in the massless limit.)
This means that the $h$-symmetry is restored {\it at} the critical
point.

Second, the fermion-scalar interactions in Eq.~(\ref{REDUCED}) are
dimension 5, and therefore irrelevant.  This (heuristically)
implies that $\theta$ decouples from the fermions near the critical
point, which is saying that the longitudinal, or gauge degrees of
freedom decouple.  The doublers are removed by the Wilson mass term,
which is present in Eq.~(\ref{REDUCED}).  The conclusion is that
a continuum limit exists (at the critical point of the reduced model) 
with free charged
left-handed fermions ({\it i.e.} fermions which couple to the 
transverse gauge field in the full theory) 
and free neutral right-handed fermions.
In other words, the fermion spectrum is chiral.  It is clear from
the discussion here that gauge-fixing plays a crucial role: without
it, the higher-derivative kinetic term for $\theta$ would not be
present.  It is the infrared behavior of this scalar field that 
causes this novel type of critical behavior to occur.  Note,
finally, that the restoration of $h$-symmetry at the critical point
and the decoupling of $\theta$ and fermion fields together imply
that the target gauge group is unbroken in the resulting continuum
theory.

Of course, the description given here is quick and dirty.  The unusual
infrared properties of this theory were investigated perturbatively
in much more detail in Ref.~\cite{wmypt}.  Fermion propagators
were computed numerically in Ref.~\cite{wmyprl}, and the agreement 
with perturbation theory was shown to be very good.  
(The numerical computations were done in
the quenched approximation.  However, the effects of quenching
occur only at higher orders in perturbation theory, so the 
good agreement between numerical and perturbative results
indicates that this is not a serious problem.)  All these 
studies confirm the results in this talk, and we refer to them
for more details.  A somewhat more extensive, but still pedagogical
account it given in Refs.~\cite{wmylat97,wmybuk}.

\section{Nonabelian speculations}

The fermionic results described in the previous section actually carry 
over to the nonabelian case.  However, in the nonabelian case, we know
that, in perturbation theory already, the ghosts do not decouple.
Omitting the ghost determinant, or something equivalent, leads to
the wrong Boltzmann weight for the gauge fields, and the resulting
theory will not be unitary.  

However, nonperturbatively, ghosts cannot be added with impunity. 
While they lead to the restoration of BRST symmetry in perturbation
theory, outside perturbation theory the existence of Gribov copies
\cite{gribov}
most likely will cause the theory to be ill-defined.  In fact, a
theorem was proven some time ago \cite{hn}, stating that, for a
BRST-invariant lattice gauge theory, the (unnormalized) expectation
value of any BRST-invariant operator vanishes identically.  The
heuristic explanation is that Gribov copies contribute to the
partition function with opposite sign for the Fadeev--Popov 
determinant, canceling their contributions in pairwise fashion.
Even in a lattice theory in which BRST symmetry is not exact on the
lattice (as is the case with our approach), one may still worry that
a similar phenomenon would occur in the continuum limit.

This particular problem would be solved if we would employ the 
absolute value of the Fadeev--Popov determinant.  However, it is not
presently known whether in that case unitarity of the theory can be
maintained at the nonperturbative level.

Here, we would like to discuss a different, ``ghost-free" approach to
gauge fixing, originally proposed in Refs.~\cite{jlp,dz}.  First
consider an exactly gauge-invariant lattice gauge theory, with
the expectation value of a gauge-invariant operator schematically
denoted by
\begin{equation}
\int {\cal D}A\;O_{\rm inv}(A){\rm exp}(-S_{\rm inv}(A))\;.
\label{OVEV}
\end{equation}
Insert unity into this expectation value in the form
\begin{equation}
1=\frac{\int {\cal D}g\;{\rm exp}(-S_{\rm ni}(A^g))}
{\int {\cal D}h\;{\rm exp}(-S_{\rm ni}(A^h))}\; \label{ONE}
\end{equation}
where the integrals are over the (compact) gauge orbit of the field $A$,
and $A^{g,h}$ are gauge transformations of $A$.  $S_{\rm ni}(A)$ is 
a non-invariant functional of the gauge field.  Changing variables
in the numerator, we find that the expectation value in Eq.~(\ref{OVEV})
is equal to
\begin{equation}
\int {\cal D}A\;O_{\rm inv}(A)
\frac{{\rm exp}(-S_{\rm inv}(A)-S_{\rm ni}(A))}
{\int {\cal D}h\;{\rm exp}(-S_{\rm ni}(A^h))}\;, \label{GFOVEV}
\end{equation} 
where we used $\int {\cal D}g=1$.  Note that in this construction 
\begin{equation}
{\rm exp}(-S_{\rm eff}(A))=
\left[\int {\cal D}h\;{\rm exp}(-S_{\rm ni}(A^h))\right]^{-1}
\label{SEFFJLP}
\end{equation}
replaces the Fadeev--Popov determinant.  

This construction has two important properties.  First, it is 
rigorously correct, in that it does not change the value of gauge-invariant
quantities.  Second, the gauge-field measure is positive definite,
which is very important from a practical point of view.

Obviously, this gauge-fixing procedure does not apply directly to
the case at hand, because the lattice action $S(A,\psi)$ from which
we start is not gauge invariant.  In order to adapt this idea to
our goal, we need to make the basic assumption that,
when gauge invariance is not preserved by the regulator, this ghost-free
approach can be made to work by adding counterterms, the form of which can be
determined in perturbation theory.  While this is an assumption,
it is not unreasonable.  First, this is exactly the way things work
in the usual perturbative formulation with ghosts, if a non-BRST-invariant
regulator is used.  Second, and more important, the existence of
continuum Gribov copies in the nonabelian case is a long-distance
property of the theory, while the need to add counterterms in order
to restore gauge invariance comes from the ultraviolet behavior of the
regulator.  Our assumption amounts to the expectation that
the form of the counterterms is not affected by 
long-distance continuum Gribov copies.

However, this does not yet mean that we have solved the problem.
Let us consider conventional perturbation theory, with ghosts.  The
action is, schematically,
\begin{equation}
S=S_{\rm classical}(A,\psi)+S_{\rm gaugefix}(A)+
\int d^4x\;{\overline c}\;\Omega(A)c+S_{\rm c.t.}(A,\psi,c,{\overline c})\;,
\label{SPERT}
\end{equation}
where $\Omega(A)$ is the Fadeev--Popov operator.  A very important 
property of the action (\ref{SPERT}) 
is that it is local.
The role of ghosts is crucial in this respect: they make
it possible to write the Fadeev-Popov determinant as a local term in
the action.  The price we have to pay is that the counterterms can also
depend on the ghost fields (and, in fact they do \cite{rossietal}).

In our ghost-free approach, the corresponding action is
\begin{equation}
S=S_{\rm classical}(A,\psi)+S_{\rm gaugefix}(A)+
S_{\rm eff}(A)+S_{\rm c.t.}(A,\psi)\;, \label{SJLP}
\end{equation}
where $S_{\rm eff}(A)$ ({\it cf.} Eq.~(\ref{SEFFJLP}))
re\-places the ghost ac\-tion, as we saw above.
However, $S_{\rm eff}(A)$ is nonlocal, and this will result in a 
nonlocal counterterm action!  As soon as the theory becomes nonlocal,
we ``loose control," and it becomes difficult, if not impossible,
to even classify the counterterms.  The problem is that $S_{\rm eff}(A)$,
while formally the same as the (logarithm of the) ghost determinant,
cannot be expressed in terms of a local action. 

This, however, does not mean that perturbation theory cannot be
formulated in a local way.  We can make some progress by examining
what happens in more detail.  For definiteness, choose a gauge-fixing
lagrangian as in Ref.~\cite{jlp},
\begin{equation}
{\cal L}_{\rm gaugefix}(A)=M^2{\rm tr}\;A^2\;,\label{LGF}
\end{equation}
with $M\sim 1/a$, and ${\rm tr}\;T^aT^b=\frac{1}{2}\delta^{ab}$, with
$T^a$ the hermitian generators of the gauge group.  
Since $M$ is not a physical mass, it is natural
to choose it of order the cutoff.  Expand $h$ in Eq.~(\ref{SEFFJLP})
as $h={\rm exp}(ig\theta/M)$, then
\begin{equation}
{\cal L}_{\rm gaugefix}(A^h)={\rm tr}\left[M^2A^2-2M\theta\,\partial_\mu A_\mu
+(\partial\theta)^2-igA_\mu[\theta,\partial_\mu\theta]\right]
+O\left(\frac{1}{M}\right)\;. \label{LGFT}
\end{equation}
Changing variables,
\begin{eqnarray}
A_\mu&=&A_\mu^T+\frac{1}{M}\partial_\mu\eta\;, \label{CHVAR} \\
\theta&\to&\theta-\eta\;, \nonumber 
\end{eqnarray}
with $\partial_\mu A^T_\mu=0$, we have
\begin{equation}
{\cal L}=\frac{1}{2}{\rm tr}\;(F_{\mu\nu}^{T})^2+M^2{\rm tr}\;A^2
+{\cal L}_{\rm fermion}(A^T,\psi)+{\cal L}_{\rm c.t.}(A^T,\eta,\psi)+
O\left(\frac{1}{M}\right) \label{NUM}
\end{equation}
in the numerator of Eq.~(\ref{GFOVEV}), and
\begin{eqnarray}
{\cal L}&=&{\rm tr}\left[M^2A^2-(\partial_\mu\eta)^2+(\partial_\mu\theta)^2
-igA^T_\mu[\theta-\eta,\partial_\mu(\theta-\eta)]\right] \label{DEN} \\
&&+{\cal L}_{\rm c.t.}(A^T,\eta,\theta)
+O\left(\frac{1}{M}\right) \nonumber
\end{eqnarray}
in the denominator.  A few important facts: the $M^2A^2$ term cancels
between numerator and denominator, showing that this is {\it not}
a (transverse) gauge-field mass term, the transverse gauge field is
massless; using shift symmetry, one can show that there are no
counterterms coupling fermions to the fields $\theta$ and $\eta$
(which both have the canonical dimension of a boson field, namely
1).  Counterterms involving
$\theta$ are, by construction, part of the denominator; we do not
have to worry about the detailed form of $O(1/M)$
terms, since they are irrelevant ($1/M\sim a$).  

We now make an important observation.  In perturbation theory, the
theory described by this (complicated) partition function, is
equivalent to the simpler theory defined by the Feynman rules
following from
\begin{eqnarray}
{\cal L}_{\rm pt}&\!=&\!\!\frac{1}{2}{\rm tr}\;(F_{\mu\nu}^{T})^2+
{\cal L}_{\rm fermion}(A^T,\psi) \label{LPT} \\
&&\!\!+\,{\rm tr}\left[(\partial_\mu\eta)^2+(\partial_\mu\theta)^2
-igA_\mu[\theta-\eta,\partial_\mu(\theta-\eta)]\right]+{\cal L}_{\rm
c.t.}+O\left(\frac{1}{M}\right)\,, \nonumber
\end{eqnarray}
with the additional rule that each connected $\theta$-subdiagram
gets an extra minus sign.  This is actually very similar to what
happens in the standard perturbative analysis with ghosts.  The
(technical) difference is that here we cannot obtain this extra
minus sign from a change in statistics of the $\theta$ field, since
not all connected $\theta$-subdiagrams are single $\theta$ loops.
($S_{\rm eff}$ is not the logarithm of a determinant.)

One more ingredient is needed.  The lagrangians of Eqs.~(\ref{NUM},%
\ref{DEN}) are nonlocal because of the presence of $A^T$, and the
decomposition (\ref{CHVAR}) is not straightforward on the lattice.
We therefore now relax the constraint that $A$ has to be transverse,
and, instead, add a new term $(1/\xi){\rm tr}\;(\partial_\mu A_\mu)^2$ to
the lagrangian, accompanied by the rule that all correlation functions
have to be calculated in the $\xi\to 0$ limit (which sets the
longitudinal part of $A$ equal to zero), before the continuum limit
is taken.  Our theory is now defined by the partition function 
(\ref{GFOVEV}) (with a functional integral over $\eta$ added), with
\begin{equation}
{\cal L}_{\rm num}=\frac{1}{2}{\rm tr}\;F_{\mu\nu}^2+
{\rm tr}\;(\partial_\mu\eta)^2
+(1/\xi){\rm tr}\;(\partial_\mu A_\mu)^2+{\cal L}_{\rm fermion}(A,\psi)
+{\cal L}_{\rm c.t.}(A,\eta,\psi) \label{NUMF}
\end{equation}
in the numerator, and
\begin{equation}
{\cal L}_{\rm den}={\rm tr}\left[(\partial_\mu\theta)^2
-igA_\mu[\theta-\eta,\partial_\mu(\theta-\eta)]\right]+{\cal L}_{\rm
c.t.}(A,\eta,\theta) \label{DENF}
\end{equation}
in the denominator, with ${\cal L}_{\rm den}$ (including
${\cal L}_{\rm c.t.}(A,\eta,\theta)$)
containing only those counterterms which
depend explicitly on $\theta$, and with the limit $\xi\to 0$
implied.  We note here that it can be proven that physical quantities
do not depend on $\xi$, and that therefore we may also define
our theory keeping $\xi$ finite.  In addition, it is possible
to show that, in perturbation theory, the partition function is
equal to the one obtained in the standard Fadeev--Popov approach
with gauge-fixing action $\int d^4x(1/\xi){\rm tr}(\partial_\mu A_\mu)^2$
\cite{prep}.

In the abelian
case the commutator term in ${\cal L}_{\rm den}$ vanishes, so that
$\theta$ decouples (just like abelian ghosts decouple in a linear
gauge).  The field $\eta$ also decouples, and the theory 
simplifies to that considered in the previous section.  (Obviously,
if $\theta$ and $\eta$ decouple, no counterterms for these fields
are required either.) 

Now we can go back to the lattice, replacing $A\to U$, $\theta\to h$,
and using Eq.~(\ref{GAUGEFIX}) in the lattice transcription of
$(1/\xi){\rm tr}\;(\partial_\mu A_\mu)^2$.  The field $\eta$ remains
a Lie-algebra valued scalar field (it is important that 
Eq.~(\ref{CHVAR}) be a linear transformation).
 
Clearly, the above described procedure needs to be checked, at the
very least to one loop in perturbation theory.  In principle,
by coupling the gauge field to a source, the Slavnov--Taylor
identities of the target theory can be constructed, and 
they can be used to construct the counterterms.  In order to
carry this out efficiently, it would be nice to have a symmetry
similar in nature to BRST symmetry in the case with ghosts, in
order to have an explicit handle on counterterms involving the
field $\theta$.

\section{Concluding remarks}

In this talk, we gave an overview of recent progress with the
gauge-fixing approach to the construction of lattice chiral gauge
theories.  The construction looks complicated, but may actually
turn out to be (relatively) practical.  There are quite a few
counterterms \cite{rome,wmky}, but (in our present formulation)
only one is of dimension 2, all others are of dimension 4.  The
dimension 2 counterterm (the gauge-field mass counterterm) will
have to be tuned nonperturbatively, but this can be done.  It is
not unlikely that, to a given typical precision, all other counterterms
can be calculated in perturbation theory to one loop (or even
tree level, {\it i.e.} omitting them altogether!).  Reference
\cite{wmky} contains more discussion of this point in the abelian
case.  In the nonabelian case, there is one more scalar field,
$\eta$, but there are not many more counterterms involving this
field, as a consequence of a shift symmetry $\eta\to\eta+$constant.
Furthermore, apart from the chiral fermion determinant, the
Boltzmann weight is positive.

An interesting feature is that we expect our approach to be
independent of the choice of lattice fermions, and, in this
sense, to be universal.  This is because the original problem
(lack of gauge invariance) is generic, as a consequence of the 
chiral anomaly.
Gauge fixing plays the central role in overcoming this problem
here, and not the choice of lattice fermions.  For encouraging
results with gauge-fixed domain-wall fermions, see Ref.~\cite{bd}.

While most of our previous work was limited to the case of
abelian chiral gauge theories, we sketched an outline
as to how the gauge-fixing approach may be extended to the
nonabelian case as well, without ignoring Gribov copies.

Another relevant issue is that of a ``spuriously" conserved
fermion number in our approach, coming from the fact that our
action is bilinear in $\psi$ and $\psibar$ \cite{banks}.
The action (and also the fermion measure) are invariant
under an exact global U(1)
symmetry which, at first glance, seems to be in contradiction
with fermion-number violation.  However, Ref.~\cite{bhs}
demonstrated, in a two-dimensional
toy model, that fermion-number violation can actually occur. 
The central observation is that fermionic states
are excitations relative to the vacuum. The global
U(1) symmetry prohibits a given state to change fermion
number, but nothing prevents the ground state
to change when an external field is applied (see also
Refs.~\cite{dm,wmypt}).
A similar phenomenon may explain how fermion-number
violating processes take place in our four-dimensional dynamical
theory.

\subsubsection*{Acknowledgements}
\medskip
We would like to thank Mike Ogilvie, Giancarlo Rossi, Andrei Slavnov, 
Jan Smit, Massimo Testa, Pierre van Baal and Arjan van der Sijs for 
discussions.  We also thank the organizers
of the workshop ``Lattice Fermions and Structure of the Vacuum"
for a very well organized and pleasant workshop.
M.G. would like to thank the Physics Departments of the
University of Rome II ``Tor Vergata," the Universitat Autonoma
of Barcelona, and the University of
Washington for hospitality.  M.G. is supported in part
by the US Department of Energy, and Y.S. is supported in part by
the Israel Science Foundation.

\end{document}